# A compact flat solar still with high performance

Guilong Peng[1], Swellam W. Sharshir[1,2,3], Rencai Ji[1], Zhixiang Hu[2], Jianqiang Ma[1], A.E. Kabeel[4], Huan Liu[2], Jianfeng Zang[2], Nuo Yang[1*]

1 State Key Laboratory of Coal Combustion, Huazhong University of Science and Technology, Wuhan 430074, China

2 School of Optical and Electronic Information, Huazhong University of Science and Technology, Wuhan 430074, China

3 Mechanical Engineering Department, Faculty of Engineering, Kafrelsheikh University, Kafrelsheikh, Egypt

4 Mechanical Power Engineering Department, Faculty of Engineering, Tanta University, Tanta, Egypt

*Corresponding email: Nuo Yang (nuo@hust.edu.cn)

# Abstract

Solar still is a convenient off-grid device for desalination, which can provide fresh water for families, ships, islands and so on. The conventional inclined solar still (ISS) suffers from low efficiency and low productivity. To improve the performance of solar still, a flat solar still (FSS) is proposed, which has a working principle similar to the solar cell. The condensate water in FSS is collected by the capillary grid attached under the ultra-hydrophilic glass cover, instead of by gravity. Therefore, FSS avoids the inclined structure and is much more compact than ISS. The daily productivity of FSS reaches up to 4.3 kg/m$^2$. Theoretical analysis shows that the enhanced mass transfer in FSS by the compact structure is an important factor for high performance. More interestingly, FSS can also be easily extended to more stage for latent heat recovery. The results show that the daily productivity of a double-stage FSS reaches up to 7 kg/m$^2$, which is much higher than the conventional solar still. FSS paves a new way in designing and optimizing of solar still.

## 1. Introduction

The fresh water is of major importance to human beings' survival as well as our economic activities such as agriculture and industry. Nowadays billions of people are suffering freshwater scarcity [1]. Thereby, it is highly desirable to find out effective ways to get fresh water from alternative resources such as wastewater, brackish, ground, sea water and so on [2]. Desalination is a promising technology to meet the global fresh water demand, due to saline water account for 97% of the water on earth's surface. Furthermore, solar desalination is considered as one type of desalination technology, which has many benefits such as no fuel cost and eco-friendly [3, 4].

The solar still, a typical small scale solar desalination system, provide a solution for the water shortage problem in remote regions, arid areas, and emergency situation etc. [5]. Solar still has many special features such as low fabrication cost, easy to maintain and portable [6]. The great majority of the investigated solar stills are inclined solar still (ISS), which mainly consists of wedge-shape basin and glass cover [7, 8]. However, solar still is not universally utilized due to the relatively low productivity as well as the low energy efficiency [9-11]. Therefore, many studies have been carried out to improve the productivity as well as the thermal performance of the solar still by different methods, such as improving the structure by designing stepped solar still [12], wick type solar still [13, 14], double slope solar still [15] and so on. Or using special materials, such as sponge [16], charcoal [17] and so forth. Nowadays, the productivity of ISS are usually 2-5 kg/m$^2$ per day and the corresponding energy efficiency is around 30-50% [7, 8, 18, 19]. Therefore, there is still ample room for further improvements.

In recent years, many efforts have been done to improve the performance of solar still by using micro/nano structured porous materials[20, 21], such as plasmonic metals [22], carbon-based materials [23], polymers [24] and semiconductors [25]. Many new materials could exhibit high solar absorptivity (>98%) across the solar spectrum [26], which indicates few energy losses during the solar absorption process. With high solar absorptivity and advanced interfacial evaporation strategy, the energy efficiency of the evaporation process can reach up to more than 90% [4, 27]. Therefore, the state-of-the-art technologies almost push the efficiency of the solar absorption process and evaporation process to the limitation.

Nevertheless, simply applying micro/nano materials to conventional solar stills might obtain a poor productivity. Many works reported that solar stills with advanced micro/nano materials are about 1-4 kg/day and the corresponding efficiency is below 40% under natural sun, which is similar to the solar still without micro/nano materials [19, 28-30]. It is found that the inefficient is resulted from the inherent disadvantages of conventional ISS, such as the inefficient condensation, solar reflection by condensate droplets, large heat loss of system, and so on [21, 29]. Therefore, there is a demand to design new types of solar stills for solving these problems and effectively improving the system performance.

Therefore, some new kinds of solar stills are studied recently, such as the thermal concentrated or thermally-localized solar still[31, 32]. In the state-of-the-art thermally-localized solar still, the evaporated water is replenished by thin hydrophilic membranes at each stage. The productivity of a 10-stage solar still can reach up to more than 2 L/(kW·h) under natural solar irradiation [33, 34]. Nevertheless, the application of such kinds of solar stills for continuously working and large-scale utilization remains a challenge, due to the salt crystallization on hydrophilic membranes or large amount of metal required for cooling.

In this paper, by imitating the working principle of solar cell, we propose a high performance flat solar still (FSS) which breaks the stereotype and makes an innovation in solar still. Similar to that electrons are collected by metallic grid in solar cell, water molecules are produced by solar energy and collected by the capillary grid in the flat solar still. FSS is capable of working continuously with neither salt crystallization on the evaporation surface nor necessary of massive metal for cooling. To illustrate the advantages of FSS, firstly, indoor experiments by using solar simulator are carried out to explore the effect of different factors on FSS. Theoretical analysis is also carried out to uncover the advantages of FSS. Lastly, the potential of constructing high efficiency double-stage FSS is verified and discussed. The proposed FSS shows great potential for large scale application and may open a new avenue for solar still system.

## 2. Experimental setup and materials

To show the difference between ISS and FSS, the schematic diagram of ISS and FSS are illustrated in Fig. 1a and 1b. In conventional ISS, the basin is wedge-shape where brackish/saline water is collected at its base. The wall of the basin in ISS is relatively high to ensure the inclined structure for water collecting. When the system works during the day time, solar irradiation entering the still through the glass and absorbed by solar absorbing materials or water. The water is heated up by solar irradiation and evaporates. The upper surface of the basin is covered by a sealed glass cover, the hot vapor flows up and condenses on the glass cover. The condensate water slides down due to gravity and accumulates outside the still as freshwater.

In FSS, the inclined structure is avoided, due to the condensate water on glass cover is collected by the capillary action of cotton threads. Thereby, the four walls of the basin in FSS have the same height and can be much lower than that in ISS (Fig. 1b) hence the areas of walls are significantly decreased and the structure is much more compact. The compact structure enhances the mass transfer between evaporation surface and the glass cover, which will be discussed later. Meanwhile, latent heat recovery is of significant importance for enhancing the daily productivity of solar still [21]. FSS can be easily extended to more stage for latent heat recovery, due to the flat and compact structure. On the contrary, most of the conventional ISS has either complex system configuration or unsatisfying performance for latent heat recovery [35-37]. Therefore, compared with ISS system, FSS has two very important advantages due to the improved system design: enhanced mass transfer and latent heat recovery performance.

The detailed schematic diagram of FSS is shown in Fig. 1c. The basin of FSS is made of foam and galvanized iron sheet, which contains saline water and prevents heat loss (Figure S3). Several foam strips float on the saline water and support the wick material on it. Saline water is transported from the basin to the black wick material (linen in this work) by capillary action, then heated by solar energy and evaporates. The floating foam and wick material enable heat localization and high efficiency evaporation at the air-water interface [38, 39]. Meanwhile, carbon black (CB) nanoparticles are dispersed on the surface of wick material to enhance solar absorption and vapor generation. The top basin is sealed by a glass cover and several cotton threads are attached to the glass cover parallelly and uniformly. When condensate water accumulates on the glass cover, it will be absorbed by the cotton threads. Later, the water absorbed by the threads will be transported out of FSS through capillary force and drop down along the vertical threads as fresh water. More details of the materials and setup can be found in Supplementary Information, Section I-III.

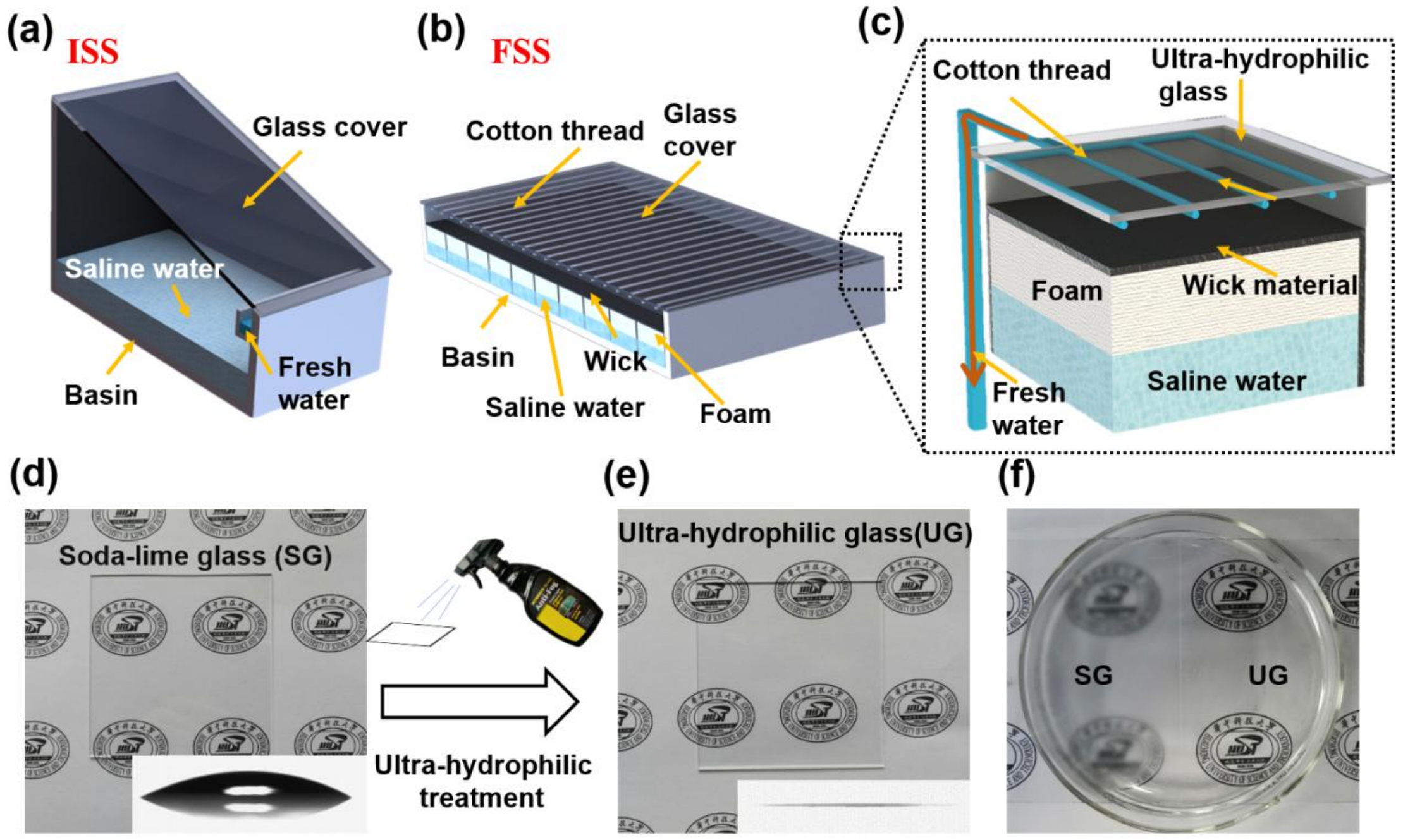

Fig.1 Setup and materials for inclined solar still (ISS) and flat solar still (FSS). The schematic diagram of (a) a conventional ISS and (b) the proposed FSS. (c) Details of the configuration of FSS. The picture of (d) conventional soda-lime glass (SG) and (e) the ultra-hydrophilic glass (UG),

the insert shows the contact angle of the glass. (f) SG and UG above hot water. Condensate water on UG forms a continuous film and is more transparent than SG.

To make FSS more efficient, the glass cover is treated to be ultra-hydrophilic. In this work, the conventional glass cover is soda-lime glass (SG) which has high transparency as shown in Fig. 1d. To make the glass ultra-hydrophilic, commercial anti-fog spray is applied to the glass surface. The contact angle before and after treatment is shown in the insert figure of Fig. 1d and 1e, which shows that the contact angle decreases significantly after treatment. The method used in this work for ultra-hydrophilic treatment is only an example. Other materials or methods might also be applied, such as using commercial anti-fog film, coating $SiO_2$ nanoparticles [40], $TiO_2$ nanoparticles [41] and $TiO_2$ nanofibers [42] et al.

The low contact angle of glass allows vapor to condense into a continuous thin film, which has two benefits. Firstly, the glass will remain clear during the condensation process, hence more solar irradiation will enter the solar still. As shown in Fig. 1f, the patterns under SG are vague due to the light diffusion and reflection by the small condensate droplets. On the contrary, the patterns under ultra-hydrophilic glass (UG) are clear and bright. Secondly, the continuous water film accelerates water transportation from glass to threads. The water film combines all the condensate water as a whole and connects with the cotton threads. Therefore, the condensate water can be absorbed by cotton threads immediately, instead of hanging under the glass as discrete droplets as happened in SG.

## 3. Results and discussions

Firstly, mini-prototypes of both FSS and ISS were made and investigated indoor by using a solar simulator. The inner volume of the small FSS is 5 cm (length) × 5 cm (width) × 3 cm (height). The intensity of solar irradiation from the solar simulator is fixed at 1kW/m$^2$ unless otherwise mentioned. The FSS with ultra-hydrophilic glass and soda-lime glass were compared by the small scale system.

The hourly water productivity of the small FSS with UG is 0.75 kg/(m$^2$·h) which is 70% higher than that of using SG (Fig. 2a). This indicates that the ultra-hydrophilic treatment of glass is very important for improving the system performance of FSS. The productivity of a small conventional ISS was also measured. It shows that there is barely fresh water collected after two hours of experiment (Fig. 2a).

Besides the effect of glass cover, the effect of cotton threads is also studied on the number of parallel thread (Fig. 2b) and the length of vertical thread (Fig. 2c). The number of thread refers to how many threads are attached to the glass cover parallelly and uniformly. The energy efficiency, η, is calculated based on the following equation [29]:

$$\eta = \frac{\Delta m \cdot h_{LV}}{A \int q\,(t) dt} \tag{1}$$

where $\Delta m$ is productivity; $h_{LV}$ is the total enthalpy of phase change, which contains latent heat and sensible heat. $h_{LV}$ can be obtained according to ref. [43]; A is the basin area of solar still; $q(t)$ is the instantaneous power density of solar irradiation, which is fixed at 1 kW/m$^2$ in the laboratory. The energy efficiency increases very slightly when the number of thread increases (Fig. 2b). The difference between 0.4 cm$^{-1}$ and 1.2 cm$^{-1}$ is less than 8%. Energy efficiency nearly converged to 54% after 0.8 cm$^{-1}$. It can be concluded that the cotton threads have very excellent water collection ability because the system performance is not greatly affected when only a few threads are used. This conclusion can be further proved by the effect of the length of vertical thread, $L_t$. The water collection is not affected even when $L_t$ is only 2 cm, which means that even very short vertical thread can provide enough driving force for water collection (Fig. 2c).

The temperatures of FSS in laboratory conditions are also measured (Fig. 2d). The temperature of both glass cover ($T_g$) and water at evaporation surface ($T_g$) raise very quickly during the first several minutes, which indicates a great solar absorption and phase change performance in FSS. After 30 minutes, the temperature is nearly converged and raises very slowly. The temperature of water at the evaporation surface and glass cover reach up to 70 °C and 65 °C , respectively, after two hours of the experiment. The maximum temperature of small FSS in laboratory is similar to that of large FSS in outdoor experiments (Figure S11).

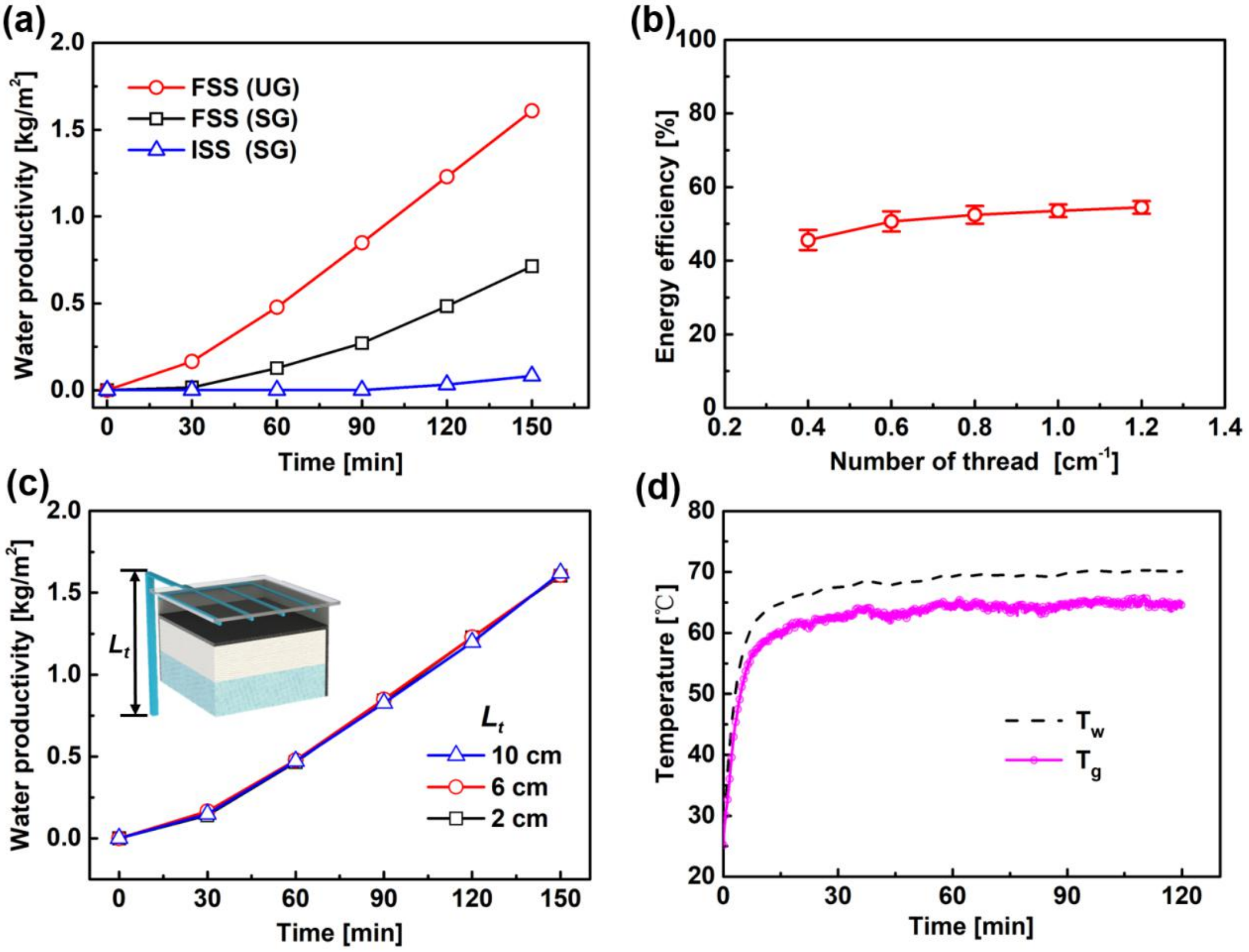

Fig. 2 Performance of a small flat solar still (FSS) and inclined solar still (ISS) in laboratory. (a) Water productivity of FSS with ultra-hydrophilic glass (UG) and conventional soda-lime glass (SG), as well as water productivity of a conventional ISS with soda-lime glass (SG). (b) The efficiency of FSS with different number of cotton threads attached to UG. (c) The productivity of FSS with different lengths of vertical cotton thread, $L_t$. (d) Temperatures of glass cover ($T_g$) and water at evaporation surface ($T_w$) in FSS. The number of cotton threads and length of vertical threads are 1 $cm^{-1}$ and 6 cm, respectively. The distance between evaporation surface and glass cover is 5 mm.

Based on the results of laboratory experiments, outdoor experiments were also carried out by homemade prototypes of FSS and ISS. Fig. 3a and 3b shows the photo of outdoor experiments and the top view of FSS. The basin area for both ISS and FSS is 25 cm × 25 cm. Water with 3.5 wt% NaCl was premixed and placed in the basin.

Bottles are used to contain the collected fresh water and the rejected brine. The solar stills and all components of the system were manufactured and tested in Shaoyang, Hunan, China (Latitude 26.99° N and longitude 111.27°E), during June to August, 2020. The daily mean ambient temperature is around 30°C, the solar irradiation various a lot due to the different weather condition. For more details about the temperature and solar irradiation of outdoor experiments, please refer to Supplementary Information, Section VIII.

In order to investigate the importance of the compact structure of FSS, the performance of FSS is compared at two different heights of vapor chamber (H), namely the distance from the evaporation surface to the glass cover. The results of a typical day (2020, August 2$^{nd}$) is illustrated in Fig. 3c, the total daily solar irradiation of this day is 5.7 kWh/m$^2$. It should be noted that the condensate water on the insulated four walls are also collected to exclude the effect of insufficient water collection. The results show that the accumulated productivity decreases around 38%, from 3.6 kg/m$^2$ to 2.6 kg/m$^2$, when H increase from 2.5 to 9.5 cm. To uncover the mechanism of the difference, theoretical model of convective mass transfer is derived based on the analogy of heat and mass transfer (Supplementary Information, Section V). The theoretical productivity agrees well with the experimental results, which indicates that the difference of productivity is mainly attributed to the different mass transfer rate in FSS. Therefore, the compact configuration of FSS is a very important advantage for improving the productivity of solar still due to the enhanced convective mass transfer.

The daily productivity of FSS and ISS under different solar irradiation is shown in Fig. 3d. The productivity of both FSS and ISS increase nearly linearly with the increase of daily solar irradiation. Around 4 kg/m$^2$ and 2.8 kg/m$^2$ of fresh water can be obtained under 6 kWh/m$^2$ of daily irradiation for FSS and ISS, respectively. It worth to be noted that the cotton threads in FSS blocks near 10% of the solar irradiation. Therefore, it can be expected that the productivity of FSS can be further enhanced by replacing the cotton threads with other transparent materials. Besides, FSS shows great salt rejecting ability and there is no salt crystallization in FSS during the experimental days, which enables FSS to work continuously for a long time. Details of salt rejecting analysis can be found in Supplementary Information, Section VI.

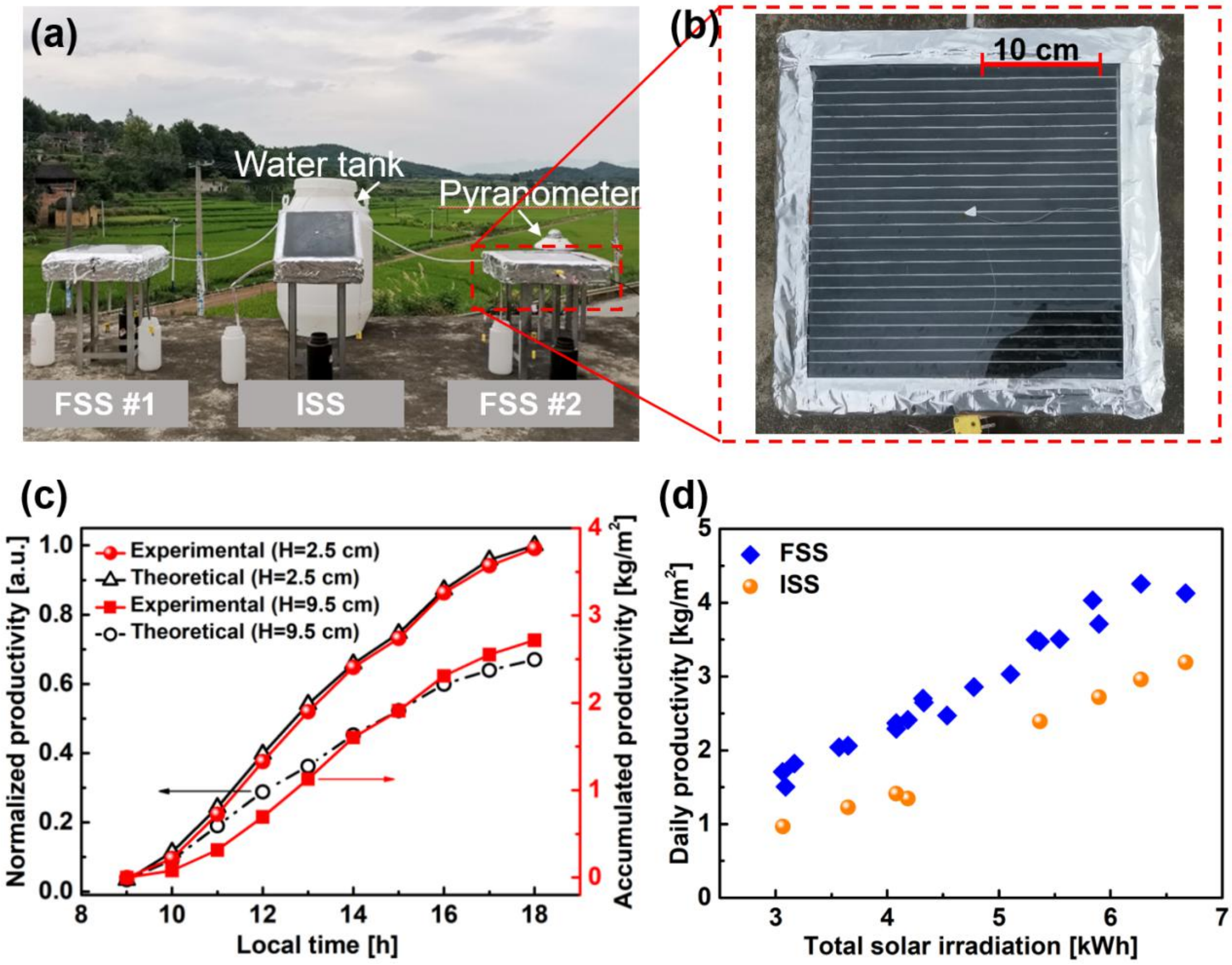

Fig. 3 Outdoor experiments and analysis for FSS. (a) The photo of ISS and FSS during outdoor experiments. (b) Top view of a FSS with lateral dimension 25 cm × 25 cm. (c) Experimental and theoretical productivity of FSS with different height of vapor chamber (H). (d) The productivity of FSS and ISS under different daily solar irradiation.

Besides the improved mass transfer due to the compact structure, FSS can also be very easily extended to more stage for latent heat recovery. Fig. 4a shows the schematic diagram a double-stage FSS. A thin saline water layer is added on the glass cover of the first stage to absorb the latent heat released by vapor condensation. The vapor generated at the second stage condenses on the glass cover above it and collected by cotton threads as in the first stage. The energy efficiency of double stage

FSS under different solar intensity is shown in Fig. 4b. It is observed that the energy efficiency of double stage FSS is sensitive to the solar intensity. When the mean solar intensity increase from 440 W/m$^2$ to 780 W/m$^2$, the energy efficiency of double stage FSS increase from 53% to 72%. This enhancement is higher than that of single stage FSS and ISS, which indicates that the superiority of a double stage FSS will be more obvious when higher solar intensity is available, such as using solar still with solar concentrator.

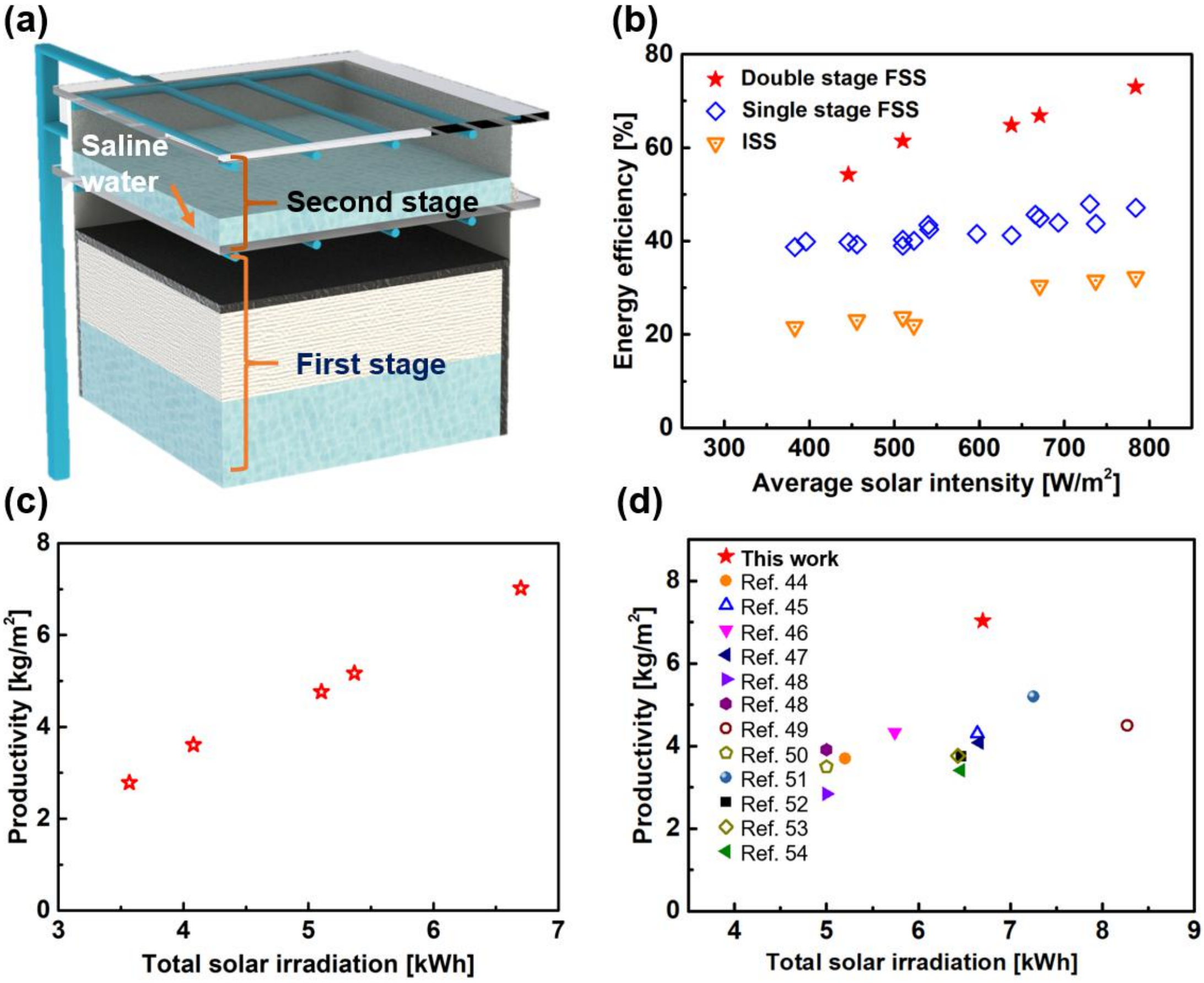

Fig. 4 Performance of the double-stage FSS. (a) The schematic diagram of double-stage FSS. (b) Energy efficiency of double stage FSS, single stage FSS and ISS under different average solar intensity. (c) Accumulated daily productivity of the double-stage FSS under different daily solar irradiation. (d) Daily productivity of solar still in this work (2 stage) and ref[44-54] (continuously working at least 8 hours, one or more stage).

The productivity also significantly increases by using double-stage FSS (Fig. 4c).

The daily productivity of double-stage FSS reaches up to 7.02 kg/m$^2$ under 6.7 kWh/m$^2$ of solar irradiation, i.e. 1.05 kg/kWh which is around 60% higher than that of single stage FSS. Higher productivity can be expected when the solar irradiation is higher during a clear day. Meanwhile, 2.8 kg/m$^2$ of fresh water can be obtained even when the solar irradiation is very weak during an overcast day (3.6 kWh/m$^2$). The high productivity of double stage FSS shows the effective and importance of latent heat recovery in solar still. Overall, the FSS proposed in this work shows great performance compared to other conventional solar stills in references. The productivity of conventional solar stills with various modification are usually between 3-5 kg/m$^2$ which is much lower than the double stage FSS as shown in Fig. 4d. It should be noted that many further modifications can be applied to FSS, such as advanced materials for solar evaporation and condensation [27, 55], solar collector or concentrator[56], glass cooling [57], and so on. The prototype of FSS in this work is only an example and better performance can be expected with more modifications.

## 4. **Conclusion**

In conclusion, based on the ultra-hydrophilic glass, cotton threads, wick materials and nanoparticles, a high efficient and compact flat solar still is proposed. The indoor results show that the wettability of glass is very important factor in FSS. The hourly productivity of FSS using ultra-hydrophilic glass is 70% higher than that of using ordinary glass. Meanwhile, the number of threads attached to the glass for fresh water collection also affect the productivity. The productivity increases a little when the number of threads increases from 0.4 to 1.2 cm$^{-1}$. Moreover, the length of thread at vertical direction doesn't affect the productivity for the range from 2 cm to 10 cm.

Furthermore, the performance of FSS is studied outdoor under natural solar irradiation. FSS could work continuously without salt crystallization on the evaporation surface. Experimental results show that the compact structure is very important for improving the productivity of FSS. Productivity decreases more than 30% when the height of vapor chamber increases from 2.5 cm to 9.5 cm. Theoretical analysis reveals that the convective mass transfer in the vapor chamber can be enhanced by the compact structure hence higher productivity. The daily productivity of single-stage FSS reaches to 4 kg/m$^2$ under 6 kWh/m$^2$ of solar irradiation. Furthermore, due to the latent heat recovery, the daily productivity of double-stage FSS reaches up to 7.02 kg/m$^2$ under 6.7 kWh/m$^2$ of solar irradiation, which is significantly higher than the conventional solar still. The high performance of FSS shows its great potential for practical application.

## 5. Conflicts of interest

There are no conflicts of interest to declare.

## 6. Acknowledgement

The work was sponsored by China Postdoctoral Science Foundation (2020M682411), National Key Research and Development Project of China (SQ2018YFE010951), National Natural Science Foundation of China (51950410592) and Fundamental Research Funds for the Central Universities (2019kfyRCPY045). The authors thank the National Supercomputing Center in Tianjin (NSCC-TJ) and China Scientific Computing Grid (ScGrid) for providing assistance in computations.

# Supplementary Information

## A compact flat solar still with high performance

Guilong Peng[1], Swellam W. Sharshir[1,2,3], Rencai Ji[1], Zhixiang Hu[2], Jianqiang Ma[1], A.E. Kabeel[4], Huan Liu[2], Jianfeng Zang[2], Nuo Yang[1*]

1 State Key Laboratory of Coal Combustion, Huazhong University of Science and Technology, Wuhan 430074, China

2 School of Optical and Electronic Information, Huazhong University of Science and Technology, Wuhan 430074, China

3 Mechanical Engineering Department, Faculty of Engineering, Kafrelsheikh University, Kafrelsheikh, Egypt

4 Mechanical Power Engineering Department, Faculty of Engineering, Tanta University, Tanta, Egypt

*Corresponding email: Nuo Yang (nuo@hust.edu.cn)

# I. Setup of laboratory experiments

The setup is shown in Fig S1. A solar simulator (CEL-S500 + AM1.5 filter) was used to generate the solar beam, the solar intensity was measured by a power meter (PM-150-50C) and adjusted to 1kW/m$^2$. FSS was insulated by foam with 3 cm of thickness. 5 mm of saline (3.5 wt.% NaCl) water layer was put inside of the FSS for desalination. The fresh water from cotton thread was collected by a small beaker. The mass of collected water was measured every half an hour by an electric balance (Sartorius Practum 224), the data were recorded by a laptop via a USB cable. The room temperature and humidity during the experiment were controlled at 25 °C and 50% respectively. The temperature was measured by T type thermal couple (Omega, TT-T-40-SLE), and a data acquisition device (Keithley 2700) were used to record the temperature.

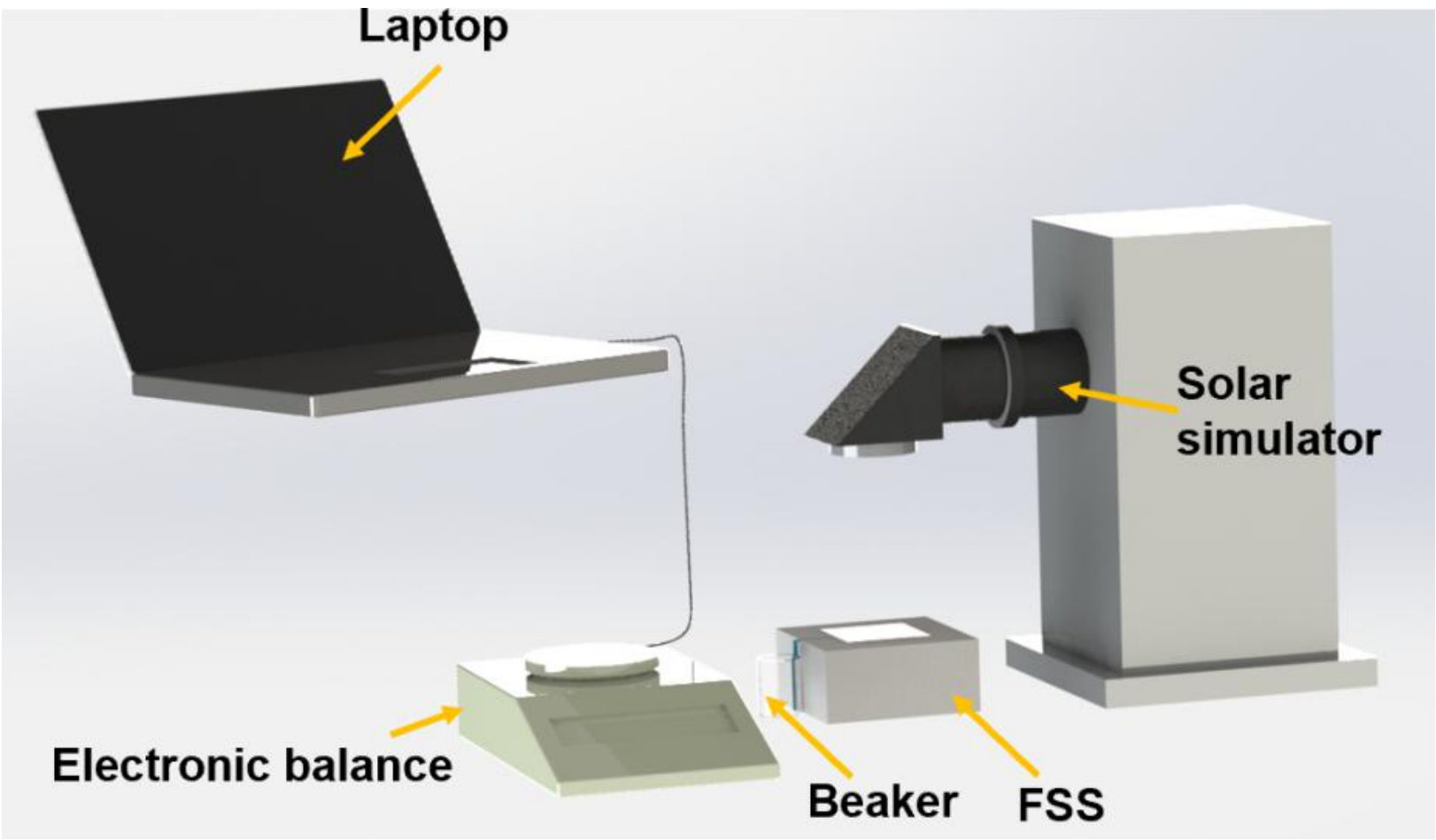

**Figure S1** The schematic diagram of measurement setup in laboratory.

**Table S1** Specification of main devices in laboratory experiments.

| Name | Type | Range | Error |
|---|---|---|---|
| Electronic balance | QUINTIX224-1CN | 0-220 g | ±0.0002g |
| Power meter | PM150-50C | 300 mW-150 W | ±3% |
| Thermal couple | TT-T-40-SLE | -200-260 °C | ±0.5°C |
| Data collector | Keithley 2700 | 1-80 Channel | $6^{1/2}$ |
| Solar simulator | CEL-S500 | 800-3000 W/m$^2$ | ±1% |

## II. Setup of outdoor experiments

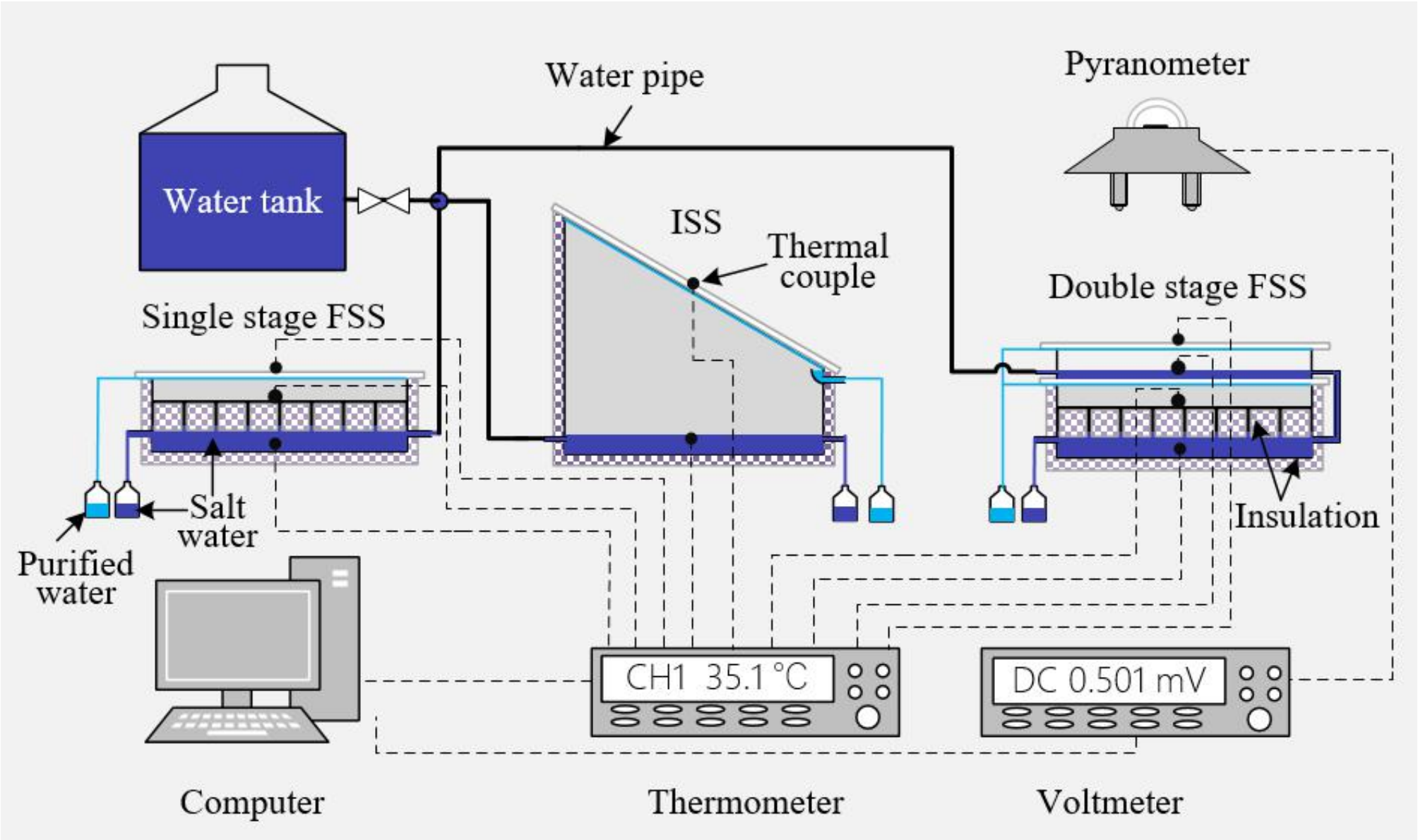

**Figure S2** The schematic diagram of measurement setup in outdoor experiments.

The setup in outdoor is shown in Figure S2. Single stage, double stage FSS and ISS are studied and compared. The water in water tank continuously flow into the solar still at a rate of around 1 kg/($m^2$·h) for water supply. The brine water is exhausted from another side of solar still and collected by bottles. The solar intensity was measured and recorded every 5 seconds by a pyranometer (TBQ-2) and a voltmeter (Keysight 34401A). The temperature of ambient and devices is measured by thermal couples (Omega TT-K-36-SLE) and thermometer (TES-1310) every hour. FSS and ISS was insulated by rubber foam with 2 cm of thickness. 10 mm of saline (3.5 wt.% NaCl) water layer was put inside of the FSS and ISS for desalination. The fresh water from cotton thread was directed by silicone hoses and collected by bottles aside the solar still. The mass of collected water was measured every hour by the electric balance.

Figure S3 shows the top view of the floating evaporation setup (FES) of FSS for outdoor experiments. FES creates heat localization on the evaporation surface for high performance solar absorption and evaporation. To effectively supply water for evaporation and rejecting the salt from the evaporation surface, the insulation foam under evaporation surface is designed as cuboid, 25 cm×3 cm×2 cm. The details of salt rejecting are discussed in Section VI. The conventional ISS that without any modification are used for comparison as many other previous works[1, 2]. The solar energy is absorbed by the black painted metallic basin in conventional ISS.

**Table S2** Specification of main devices in outdoor experiments.

| Name | Type | Range | Error |
|---|---|---|---|
| Electronic balance | K-FINE 500 | 0-500 g | ±0.01g |
| Power meter | TBQ-2 | 0 -2000 W/m$^2$ | ±2% |
| Thermal couple | Omega TT-K-36-SLE | -267-260 °C | ±0.5°C |
| Thermometer | TES-1310 | -50-1300°C | ±0.1°C |
| Voltmeter | Keysight 34401A | 0-1000 V | ±0.005% |

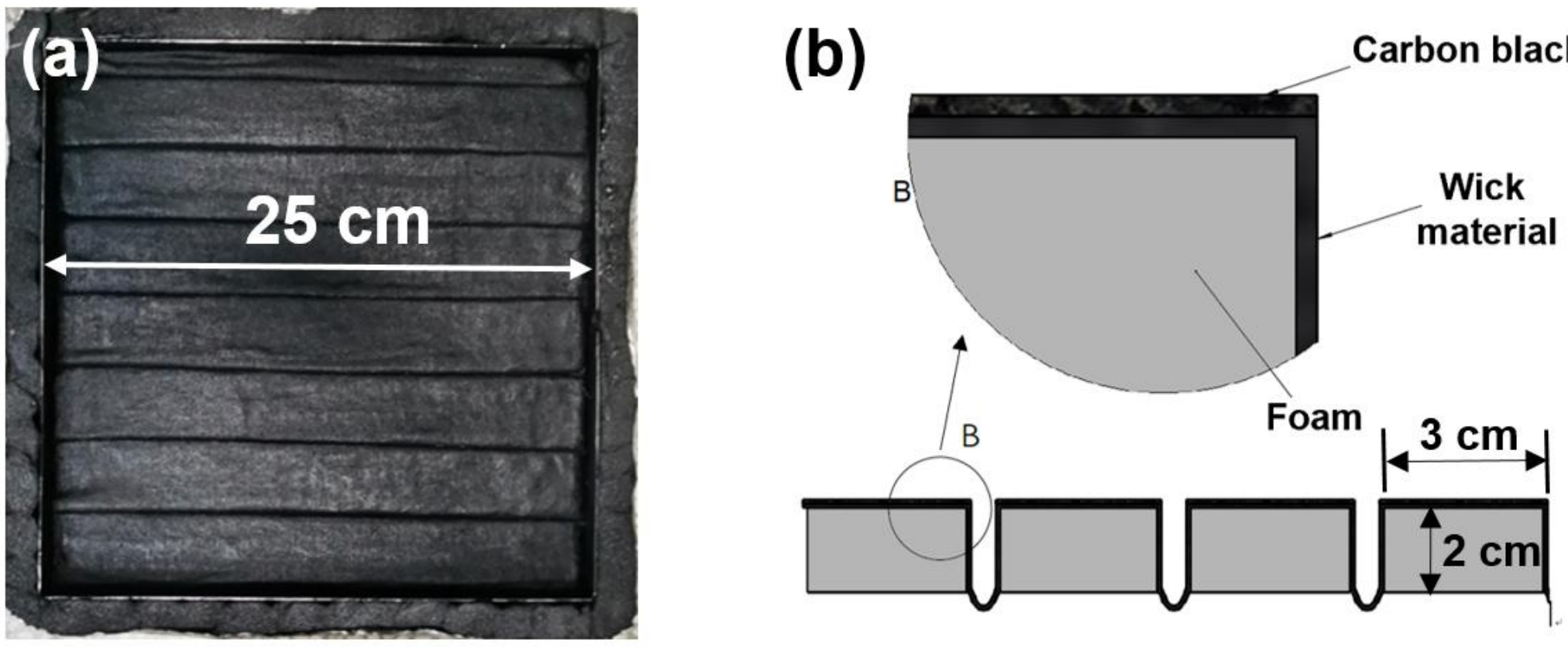

**Figure S3** (a) Top view of the floating evaporation setup (FES) of FSS. (b) Schematic diagram of the side view of the FES in FSS.

# III. Wick material and cotton threads

In this work, the black linen cloth and cotton thread were chosen as the wick material, (shown in Figure S4 and Figure S5). However, some other daily materials can also be chosen as the wick material, such as tissue, paper and so on, as long as the capillary action is strong enough [3]. The capillary action ability of the wick material can be described by the rate of moisture regain, which is 12.5% for linen [4] and 8.5% for cotton [5]. A double layer structure of linen cloth is used in FSS for water evaporation, which enables great water absorption ability, hence preventing salt accumulation on evaporation surface [6]. The mass density of carbon black (CB) nanoparticles on the surface of wick material is 10 g/m$^2$.

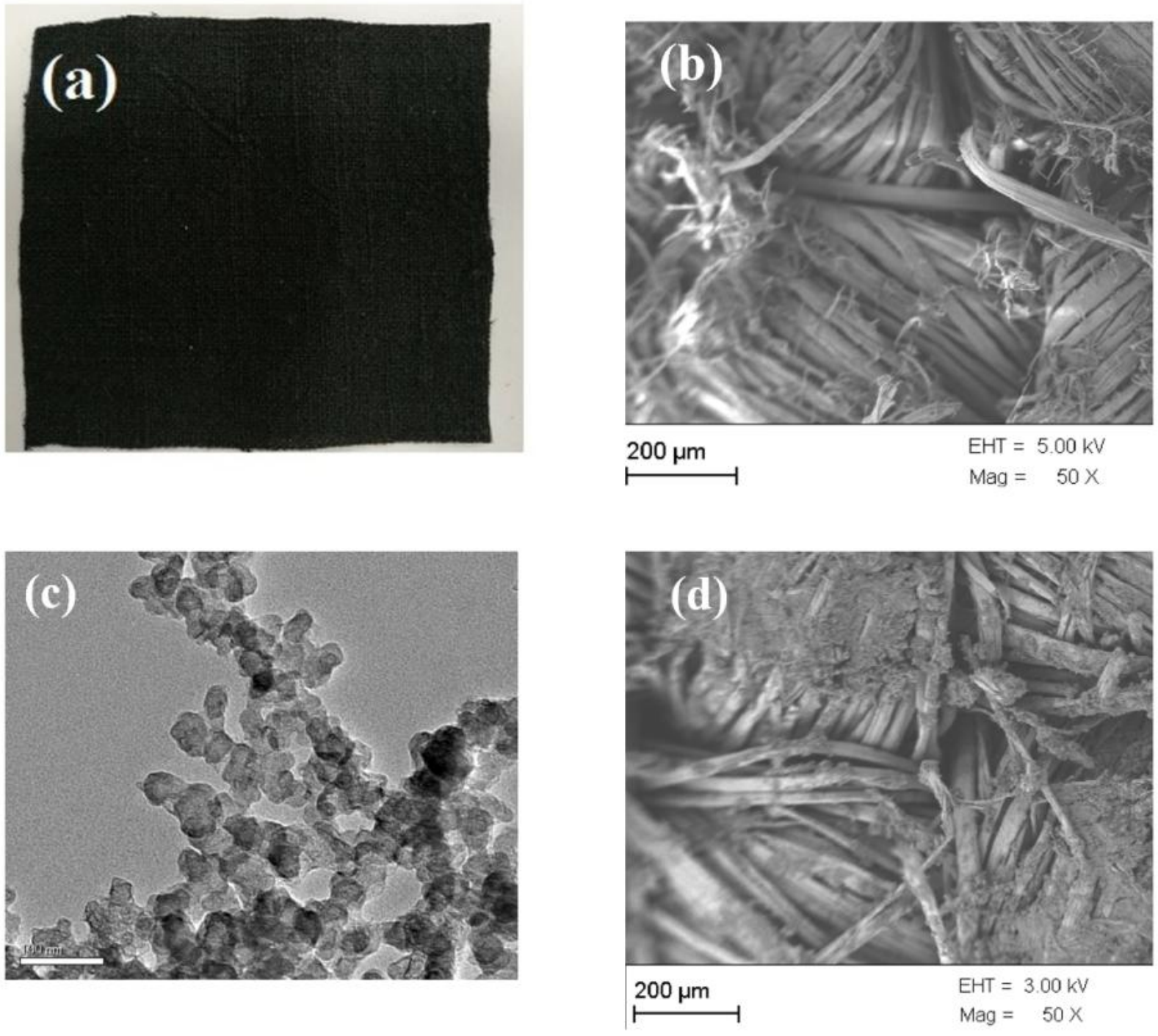

**Figure S4** (a) The picture of linen cloth. (b) The scanning electron microscope (SEM) image of linen cloth. (c) The transmission electron microscope (TEM) of carbon black (CB) nanoparticles. The size of CB is around 30-40 nm. (d) The SEM image of CB on linen cloth.

**Table S3** The characteristics of the wick material and carbon black.

| **Material** | **Characteristic** | Value |
|---|---|---|
| Linen | Rate of moisture regain [%] | 12.5 |
| | Mass density [g/m$^2$] | 250 |
| | Thickness [mm] | 0.5 |
| Cotton | Rate of moisture regain [%] | 8 |
| | Diameter [mm] | 0.4 |
| Carbon black | Diameter [nm] | 30-40 |

In laboratory experiment, the thread under the glass only contains a single cotton thread as shown in Figure S5 (a). However, 5 single cotton threads are twisted into a bundle as shown in Figure S5(b), and attached under the glass cover in outdoor FSS. This is because that the length of water film along the transporting direction in outdoor FSS is 5 times that of the small FSS in laboratory, hence 5 times larger threads are needed.

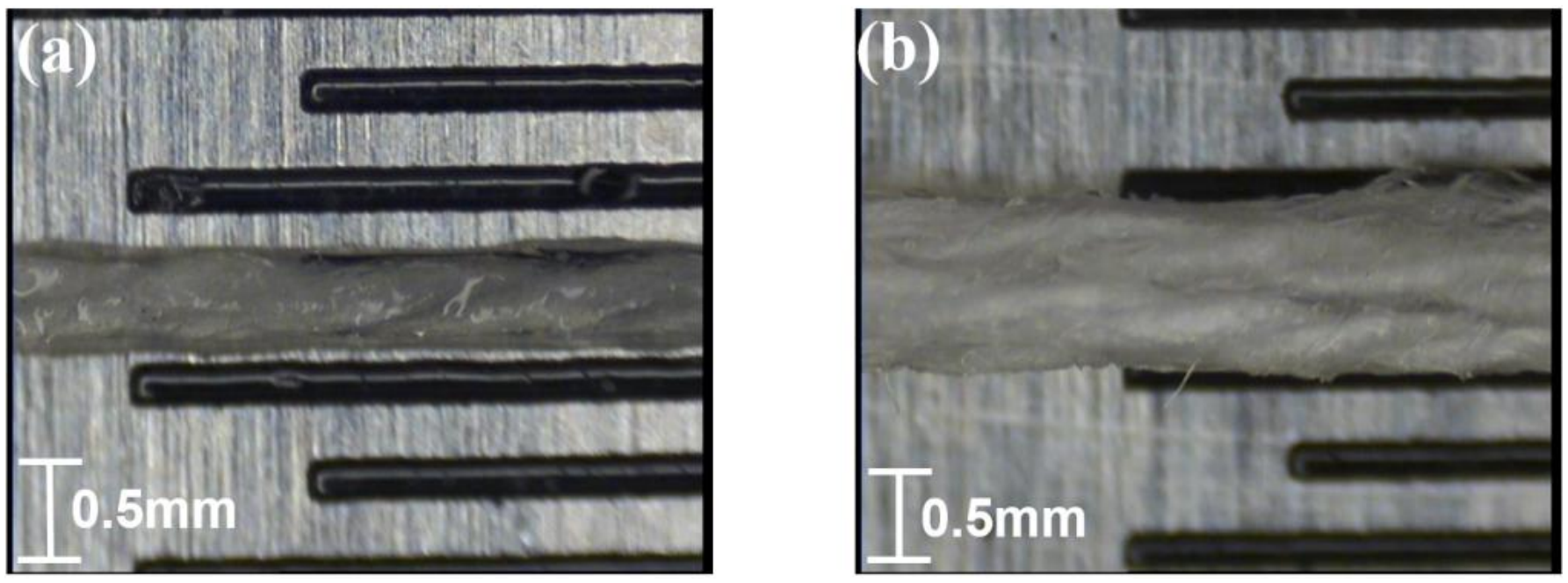

**Figure S5** (a) A single wetted cotton thread. The diameter of a single cotton thread is around 0.4 mm. (b) A bundle of wetted cotton threads that consist of 5 single cotton threads, the diameter is around 1 mm.

## IV. Solar absorption and evaporation of FSS

The solar absorptivity of black wick material (linen) and wick material with carbon black is measured in our previous study[7]. Black wick material has very high absorptivity (around 99%) in the visible light region, the absorptivity later decreases to 80% in the near-infrared region. On the other hand, with the help of carbon black, the absorptivity keeps at 97% in almost the entire solar spectrum.

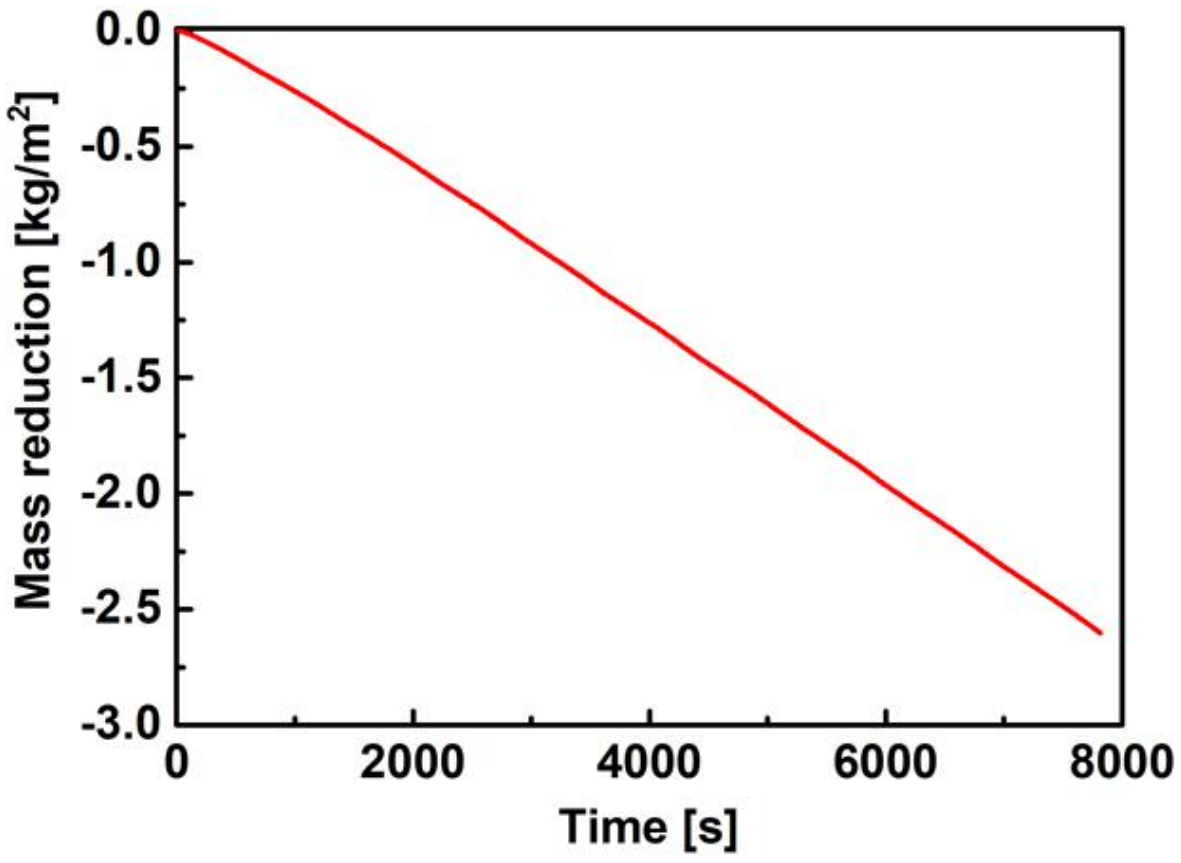

**Figure S6** The solar evaporation performance of the small scale FSS.

The evaporation performance of the small scale FSS ( 5 cm × 5 cm × 3 cm ) is measured as shown in Fig. S6. During the measurement process, the glass cover and cotton threads of FSS are removed. Water evaporates from the surface of wick material, and the generated vapor diffuses freely to the environment, which is the same as the works for solar evaporation[8, 9]. The total evaporation rate is around 1.25 kg/($m^2$·h) under 1 kW/$m^2$ of solar intensity. The evaporation rate from references can reach up to 1.5~4 kg/($m^2$·h) by using advanced nano/micro materials[10-12]. Therefore, the productivity of FSS can be further increased by integrating with advanced solar evaporation technologies.

# V. Convective mass transfer model

The productivity, $\dot{m}$, of FSS is the amount of condensed water on the glass cover, which is decided by the condition of both evaporation and condensation surface, as well as the structure of the phase change chamber. Based on heat and mass analogy, $\dot{m}$ can be obtained by[13]:

$$\dot{m} = h_m A\left(m_w^{'} - m_g^{'}\right) \qquad \text{(S1)}$$

where $h_m$ is the convective mass transfer coefficient between evaporation surface and glass cover, A is the area of FSS, $m_w^{'}$ and $m_g^{'}$ are the mass fractions of water vapor in the saturated air at the evaporation surface and the glass cover, respectively.

The mass fraction is related to the saturated vapor pressure, $P_s$, and the total pressure, $P$:

$$m' = \frac{P_s}{P}\rho_a \tag{S2}$$

where $\rho_a$ is the density of the moist air, the saturated vapor pressure is determined by the moist air temperature $T_{ma}$[7]:

$$P_s = e^{\left(25.317-\frac{5144}{T_{ma}}\right)} \tag{S3}$$

The total pressure P in the vapor chamber is regarded as the ambient air pressure, due to the chamber is connected to the ambient through water feeding and rejecting pipes.

The convective mass transfer coefficient $h_m$ can be obtained by using the heat and mass transfer analogy. For heat transfer in a small horizontal chamber with hot surface at the bottom, the Nusselt number is described as [14]:

$$Nu = 0.212(GrPr)^{\frac{1}{4}} \tag{S4}$$

where $Gr$ and $Pr$ are the Grashof number and Prandtl number, respectively. $Gr$ and $Pr$ are defined as:

$$Gr = \frac{g\alpha_v \Delta T \delta^3}{\nu^2} \tag{S5}$$

$$Pr = \frac{\nu}{\alpha} \tag{S6}$$

where $g$ is the gravitational acceleration, $\alpha_v = 1/T_{ma}$ is the volume expansion coefficient of moist air and $\alpha$ is the thermal diffusivity of moist air, $\delta$ is the characteristic size of chamber, which equals to the height of FSS, i.e. $\delta = H$. $\Delta T$ is temperature difference between the evaporation surface and the glass cover, $\nu$ is the kinematic viscosity of moist air.

Based on the heat and mass transfer analogy, we have [13]:

$$\frac{h_h}{h_m} = \rho_a C p_a L e^{2/3} \tag{S7}$$

where $h_h$ is the convective heat transfer coefficient, $Le$ is the Lewis number:

$$h_h = \frac{Nu\lambda}{\delta} \tag{S8}$$

$$\mathrm{Le} = \frac{\alpha}{D} \tag{S9}$$

where λ is the thermal conductivity of air, D is the diffusion coefficient of water vapor in air:

$$D = D_{298K}\left(\frac{T_{ma}}{298}\right)^{1.5} \qquad (S10)$$

where $D_{298K} = 0.256 \times 10^{-4} \mathrm{m}^2/\mathrm{s}$ is the diffusion coefficient at 298K.

The thermophysical property of the saturated moist air for calculation can be obtained by the following equations [15]:

$$\alpha = \mathrm{SA}_0 + \mathrm{SA}_1 t + \mathrm{SA}_2 t^2 + \mathrm{SA}_2 t^3 + \mathrm{SA}_4 t^4 \qquad (S11)$$

$$v = \mathrm{SV}_0 + \mathrm{SV}_1 t + \mathrm{SV}_2 t^2 + \mathrm{SV}_2 t^3 + \mathrm{SV}_4 t^4 \qquad (S12)$$

$$\lambda = \mathrm{SK}_0 + \mathrm{SK}_1 t + \mathrm{SK}_2 t^2 + \mathrm{SK}_2 t^3 + \mathrm{SK}_4 t^4 \qquad (S13)$$

$$\rho = \mathrm{SD}_0 + \mathrm{SD}_1 t + \mathrm{SD}_2 t^2 + \mathrm{SD}_2 t^3 \qquad (S14)$$

The value of coefficients in the equations are listed in Table S4, *t* is average temperature of the moist air in degree Celsius. Based on Eq. S1-S14 and the temperature of glass cover and the evaporation surface measured by experiment, the theoretical productivity of FSS can be obtained. The measured temperature is shown in Fig. S11.

Table S4 List of coefficients for calculating the thermophysical property of the saturated moist air (0 – 100 °C) [15].

| | 0 | 1 | 2 | 3 | 4 |
|---|---|---|---|---|---|
| SA | 1.847E-5 | 1.162E-7 | 2.373E-10 | -5.769E-12 | -6.369E-14 |
| SV | 1.716E-5 | 4.722E-8 | -3.663E-10 | 1.873E-12 | -8.050E-14 |
| SK | 24.007E-3 | 7.278 E-5 | -1.788E-7 | -1.352E-9 | -3.322E-11 |
| SD | 1.293 | -5.538E-3 | 3.860E-5 | -5.254E-7 | – |

# VI. Salt rejecting model

The salt rejecting ability of solar still is very important for preventing salt crystallization. The traditional solar still directly heats up the bulk water to evaporate without the problem of salt crystallization. However, the evaporation setup in this works contains very thin layer of water. Therefore, if the evaporation is too fast, the salt concentration will be too high on the evaporation surface, and finally lead to salt crystallization. In order to guide the design of evaporation setup, the effect of the structure parameters of the evaporation setup should be discussed.

Fig. S7 shows the schematic diagram of the cross section of an evaporation unit. Given that the evaporation unit does not change along the Y direction (perpendicular to paper), only x-z section is taken as the analysis object. There are three main structural parameters that affect the salt rejecting, including the half width of evaporation surface $L_1$, the height of evaporation surface $L_2$ and the thickness of wick material $\delta_f$. During the evaporation process, a salt concentration gradient is formed along the direction of salt diffusion (Direction X). The concentration of salt at three positions are important in analysis, including the concentration of salt in the center of the evaporation surface $C_1$, the concentration of salt at the edge of the evaporation surface $C_2$ and the initial concentration of salt water $C_3$. In analysis, the salt concentration gradient in the direction of thickness of wick material is ignored.

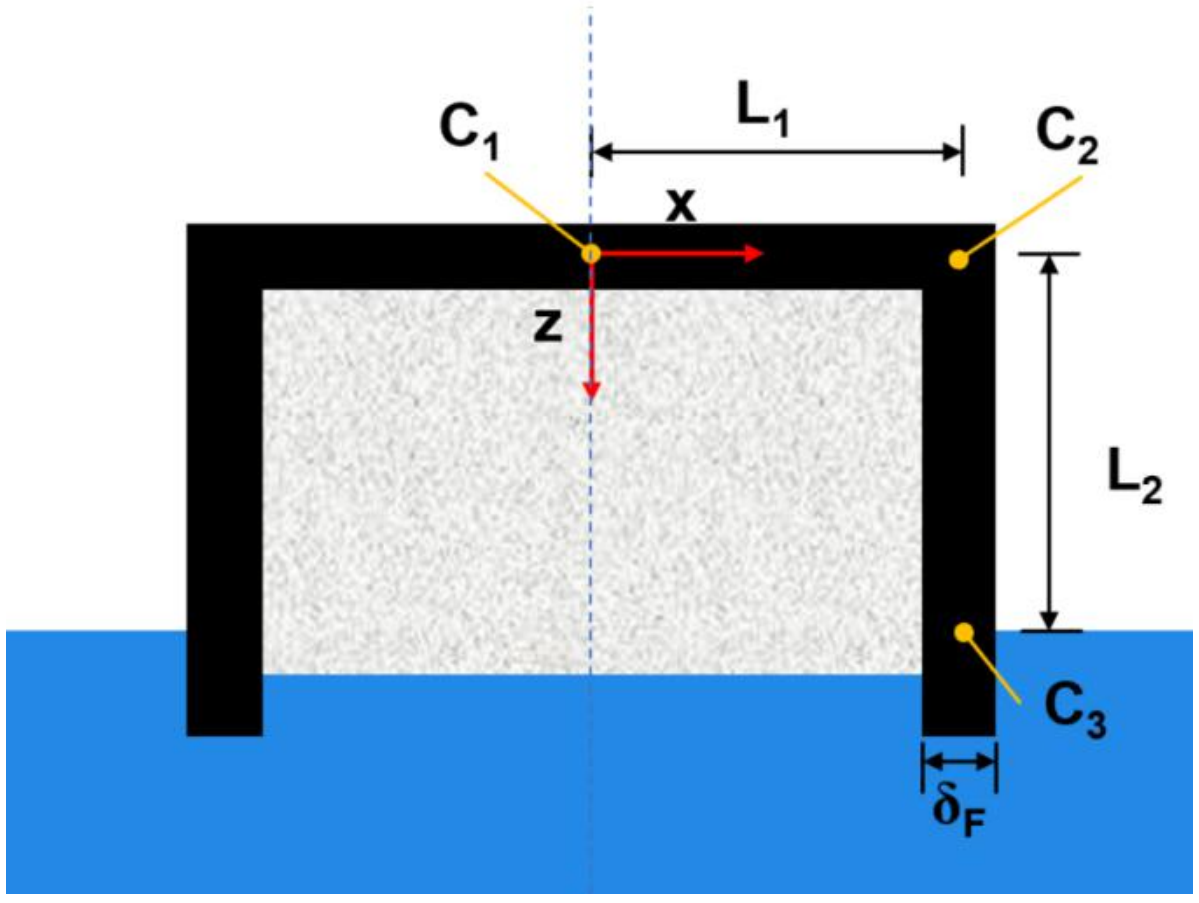

**Figure S7** Schematic diagram of the cross section of a evaporation unit

The salt rejecting process in the wick material can be described as:

$$J = D_{NaCl}\delta_F\rho\frac{\Delta C}{L} \quad \text{(S15)}$$

Where $J$ is the mass flux of salt, $D_{Nacl}$ is the diffusion coefficient of NaCl in water. When there is no advection, $D_{NaCl} = 1.99 \times 10^{-9}\ m^2/s$. ρ is the density of brine water (~1.3 g/cm$^3$), $\Delta C$ is the difference of salt concentration, $L$ is the diffusion length.

In the evaporation process, the mass of salt accumulated on the evaporation surface for a unit length at y direction is:

$$M_{NaCl}^{E}=\frac{C_3}{1-C_3}\left(L_1+\frac{\delta_F}{2}\right)\dot{m} \quad \text{(S16)}$$

Where $\dot{m}$ is the evaporation rate of water per unit area.

When the evaporation is stable, the salt rejected by the wick material is equal to the salt contained in the evaporated water. In the vertical direction of the wick material, evaporation does not occur. Therefore, all salt absorbed by the wick material reaches the horizontal evaporation region through vertical non evaporation region, and the accumulated salt in evaporation region is also rejected into the bulk water through non evaporation region. Thereby:

$$\frac{C_3}{1-C_3}\left(L_1+\frac{\delta_F}{2}\right)\dot{m}-D_{NaCl}\delta_F\rho\frac{C_2-C_3}{L_2}=0 \quad \text{(S17)}$$

In evaporation region, the water transportation and salt diffusion are shown in Fig. S8. The mass of salt produced ($M_{NaCl}^{E}(x)$)and rejected ($J(x)$) at each $dx$ are:

$$M_{NaCl}^{E}(x)=\dot{m}_x\frac{C_x}{1-C_x}dx \quad \text{(S18)}$$

$$J(x)=D_{NaCl}\delta_F\rho\frac{dC_x}{dx} \quad \text{(S19)}$$

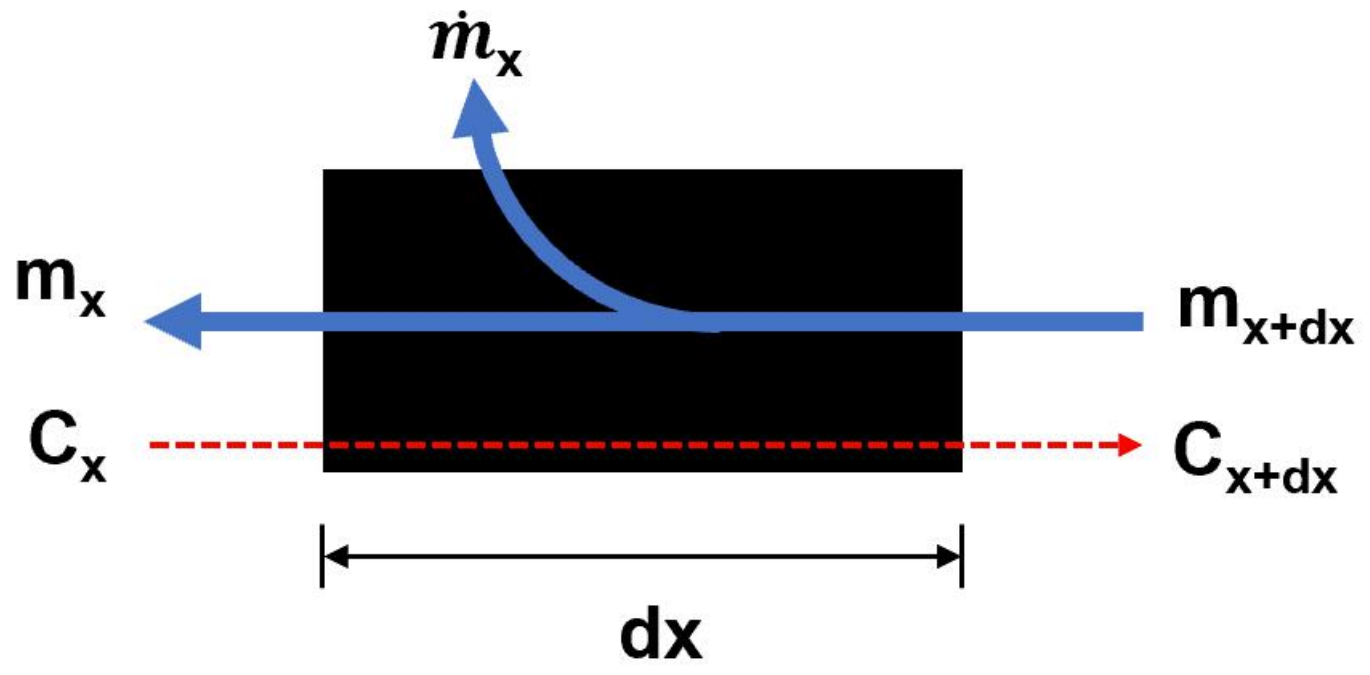

**Figure S8** Mass conservation of water and salt.

Based on the mass conservation of water and salt, we can obtain that:

$$-D_{NaCl}\delta_F\rho\frac{dC_x}{dx}-\int_0^x\dot{m}_x\frac{C_x}{1-C_x}dx=0 \quad \text{(S20)}$$

$$\int_0^{L_1+\delta_F/2}\dot{m}_x\,dx=\left(L_1+\frac{\delta_F}{2}\right)\dot{m} \quad \text{(S21)}$$

$$\int_{0}^{L_1 + \delta_F/2} \dot{m}_x \frac{C_x}{1-C_x} dx = \frac{C_3}{1-C_3}\left(L_1 + \frac{\delta_F}{2}\right)\dot{m} \qquad (S22)$$

Assume that $\dot{m}_x$ is constant and equals to $\dot{m}$, than:

$$-D_{NaCl}\delta_F\rho\frac{dC_x}{dx} - \int_0^x \dot{m}\frac{C_x}{1-C_x}dx = 0 \qquad (S23)$$

Therefore:

$$-D_{NaCl}\delta_F\rho\frac{d^2C_x}{dx^2} - \dot{m}\frac{C_x}{1-C_x} = 0 \qquad (S24)$$

The boundary conditions is:

$$x + dx = L_1 + \frac{\delta_F}{2} \qquad C_{x=L1+\frac{\delta_F}{2}} = \frac{\frac{C_3}{1-C_3}\left(L_1 + \frac{\delta_F}{2}\right)\dot{m}L_2}{D_{NaCl}\delta_F\rho} + C_3 \quad (S25)$$

When salt crystallizes at the center, $C_1$=26 wt%, therefore:

$$x = 0 \qquad C_{x=0} = 0.26 \qquad (S26)$$

$$C'_{x=0} = 0 \qquad (S27)$$

Combining Eq. S24-S27, S24 can be solved by the fourth-order Runge Kutta method.

In order to obtain the actual salt diffusion coefficient in the experiment, we measured the salt crystallization conditions. The experimental results show that when the evaporation rate $\dot{m}$ is 1.2 kg/(m$^2$·h), the half width of the evaporation surface $L_1$ = 1.5 cm, the height of the evaporation surface $L_2$ = 2 cm, and the initial concentration of salt water $C_3$ = 3.5 wt%, salt will crystallize at the center if $\delta_f$ = 0.5 mm (single-layer linen). However, if $\delta_f$ = 1 mm (double-layer linen), there is no salt crystal at the center of the evaporation surface. Therefore, assuming that $\delta_f$ = 0.75mm is the critical point of salt crystallization, the actual salt diffusion coefficient is $D_{Nacl} \approx 6.1 \times 10^{-8}\ m^2/s$. The calculated actual diffusion coefficient is much larger than that of salt when the water is still, which shows that salt can be rejected to the bulk water through advection under the effect of temperature gradient, concentration gradient, gravity and capillary flow.

The salt rejecting performance of the evaporation setup is further verified in the outdoor experiments as shown in Figure S9. The glass cover of FSS is removed and FSS is placed outdoor under sunshine for solar evaporation. Water with 3.5%wt. of NaCl is placed under the evaporation surface. To better show the salt rejecting performance, extra salt powder is sprayed on the evaporation surface. It is observed that the salt powder is diminishing during the evaporation process, which indicates that the evaporation setup has enough salt rejecting ability for practical application.

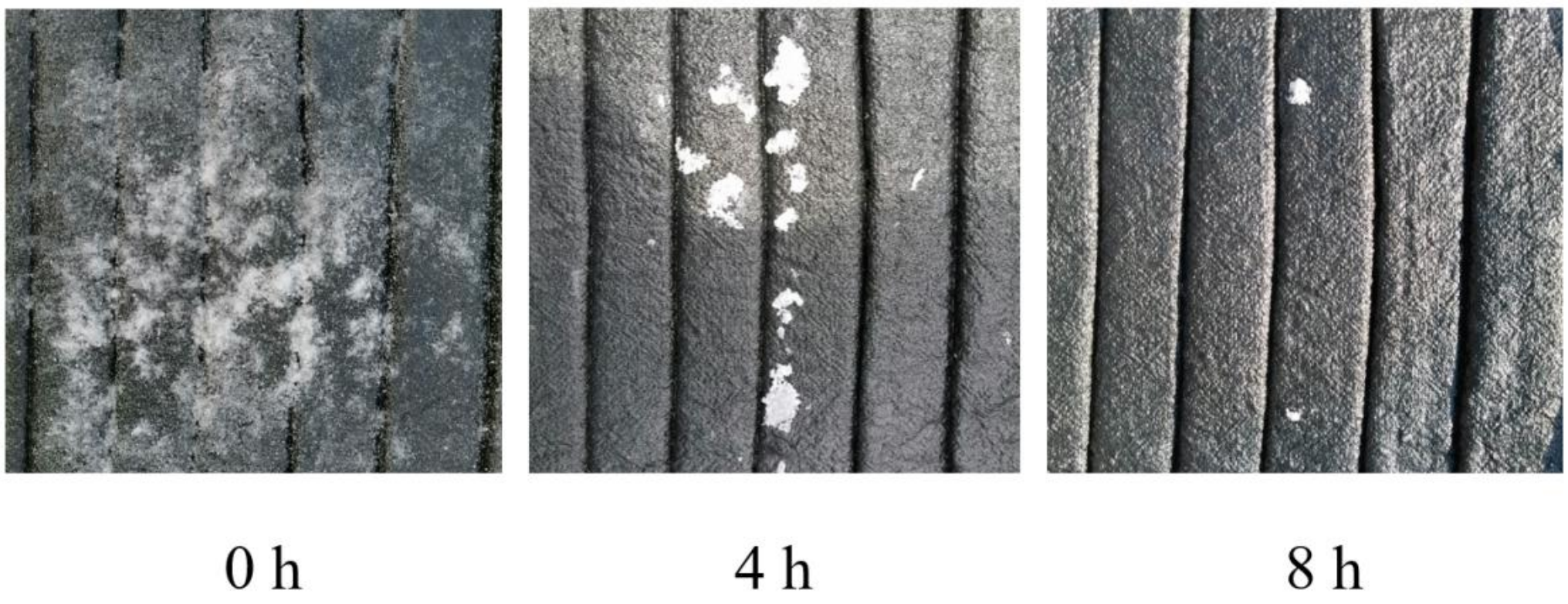

**Figure S9** Salt rejecting during the evaporation process.

## VII. Comparison between modified FSS and modified ISS

The comparison between modified FSS and modified ISS is also studied in order to reveal the importance of each modification step by step (Figure S10). The results indicate that the efficiency of conventional ISS with ultra-hydrophilic glass (ISS(UG)) is higher than that of ISS with ordinary glass (ISS(SG)) by around 4-7%. Therefore, the better transmittance of glass cover resulted from ultra-hydrophilic treatment is helpful to the improvement. Another important point is FES, which creates heat localization on the evaporation surface and enhances the efficiency of FSS and by around 10-20%. The double stage FSS with ultra-hydrophilic glass and floating evaporation setup (2 stage FSS(UG+FES)) further improves the efficiency by around 20%. The efficiency of different types of solar stills under a similar solar irradiation (~520 W/$m^2$) is listed in Table S5, which shows the enhancement clearly.

It should be noted that the difference between ISS and FSS should be more obvious if transparent threads are used to avoid blocking the solar irradiation. The cotton threads account for around 10% of the total area hence the input solar irradiation in FSS is 10% less than that in ISS. Besides, the condensation area in FSS is smaller than that of ISS by 15%, which decrease the condensation rate of FSS as compared to ISS. This disadvantage can be avoided if FSS is used in active mode by using external condensation.

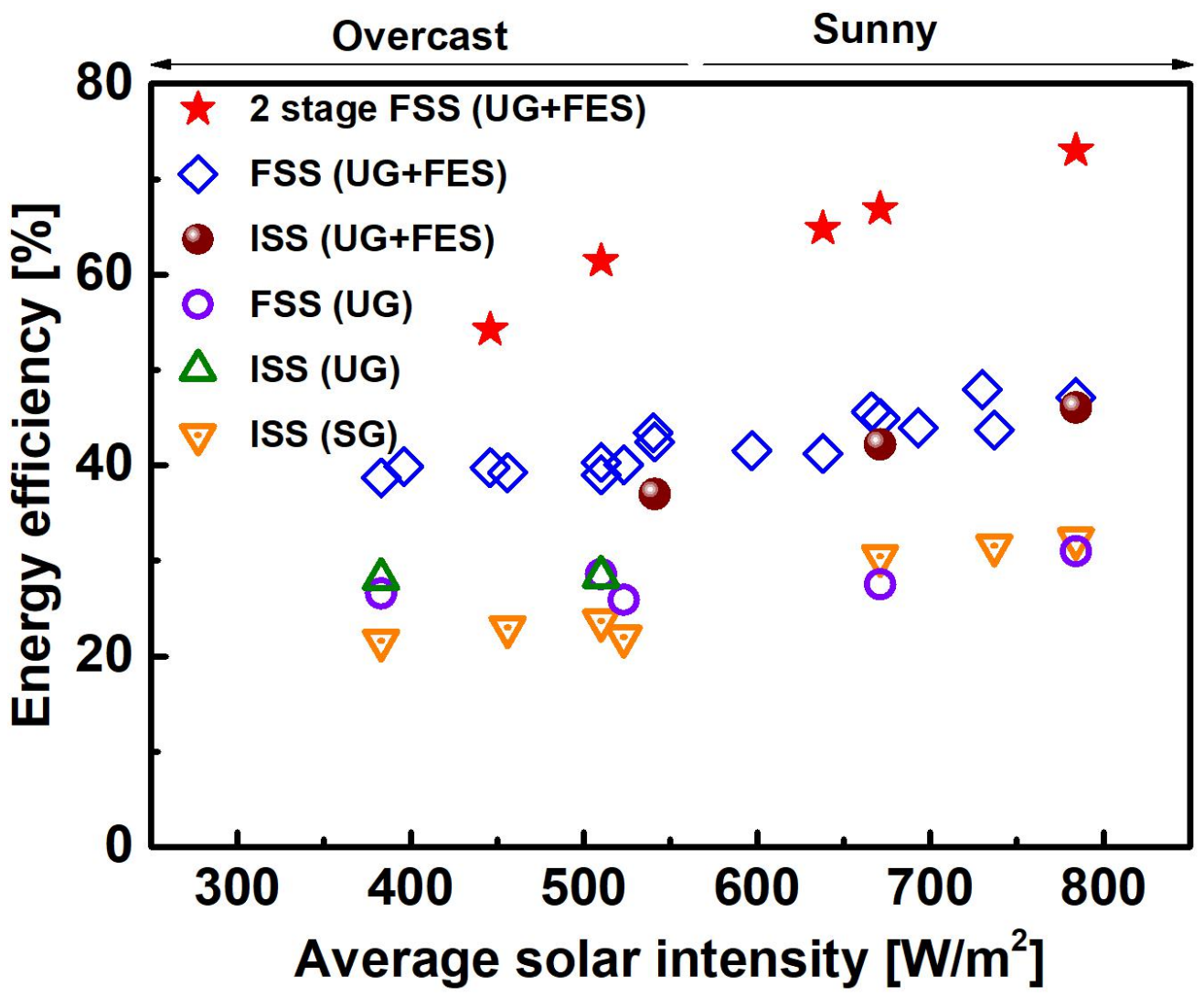

**Figure S10** Energy efficiency of different types and modifications of solar still.

**Table S5** Efficiency and enhancement of different types of solar stills under ~520 W/m$^2$ of solar irradiation.

| Type | Efficiency (%) | Enhancement compared to ISS(UG) (%) |
|---|---|---|
| ISS(SG) | 24.4 | 0 |
| ISS(UG) | 28.8 | 18% |
| FSS(UG) | 28.6 | 17% |
| ISS(UG+FES) | 37.1 | 52% |
| FSS(UG+FES) | 41.6 | 70% |
| 2 stage FSS(UG+FES) | 60.2 | 147% |

## VIII. Temperature and solar irradiation of outdoor experiments

The outdoor experiments were carried out for many days. Herein, the temperature and solar irradiation of five typical days are presented. The ambient temperature during the experiments is around 30±3 °C as shown in Figure S11(a). The temperature of water and glass cover of FSS varies a lot with the different solar irradiation on different days. The maximum temperature of water and glass cover in FSS reaches up to more than 75 °C and 67 °C, respectively (Figure S11(b) and (c)). In the double stage FSS, the temperature of water at the first stage ($T_{w1}$), the second stage ($T_{w2}$) and the glass cover at the second stage ($T_{g2}$) can reach up to around 80°C, 75 °C and 70°C, respectively. The temperature of FSS with different height of vapor chamber (H) is shown in Figure S11(d). The water temperature of H=9.5 cm is obviously lower than that of H=2.5 cm by 1-5 °C.

The solar irradiation of five days is presented in Figure S12(a)-(e). The solar intensity fluctuates a lot due to the clouds. The maximum solar intensity reaches up to more than 1400 W/m$^2$ (Figure S12 (e)) due to the strong scattered radiation by clouds. When the sky is more clear, the maximum solar irradiation is around 1000 W/m$^2$ at noon (Figure S12 (d)). The wind velocity during the experiments is less than 4 m/s, and even windless on some days.

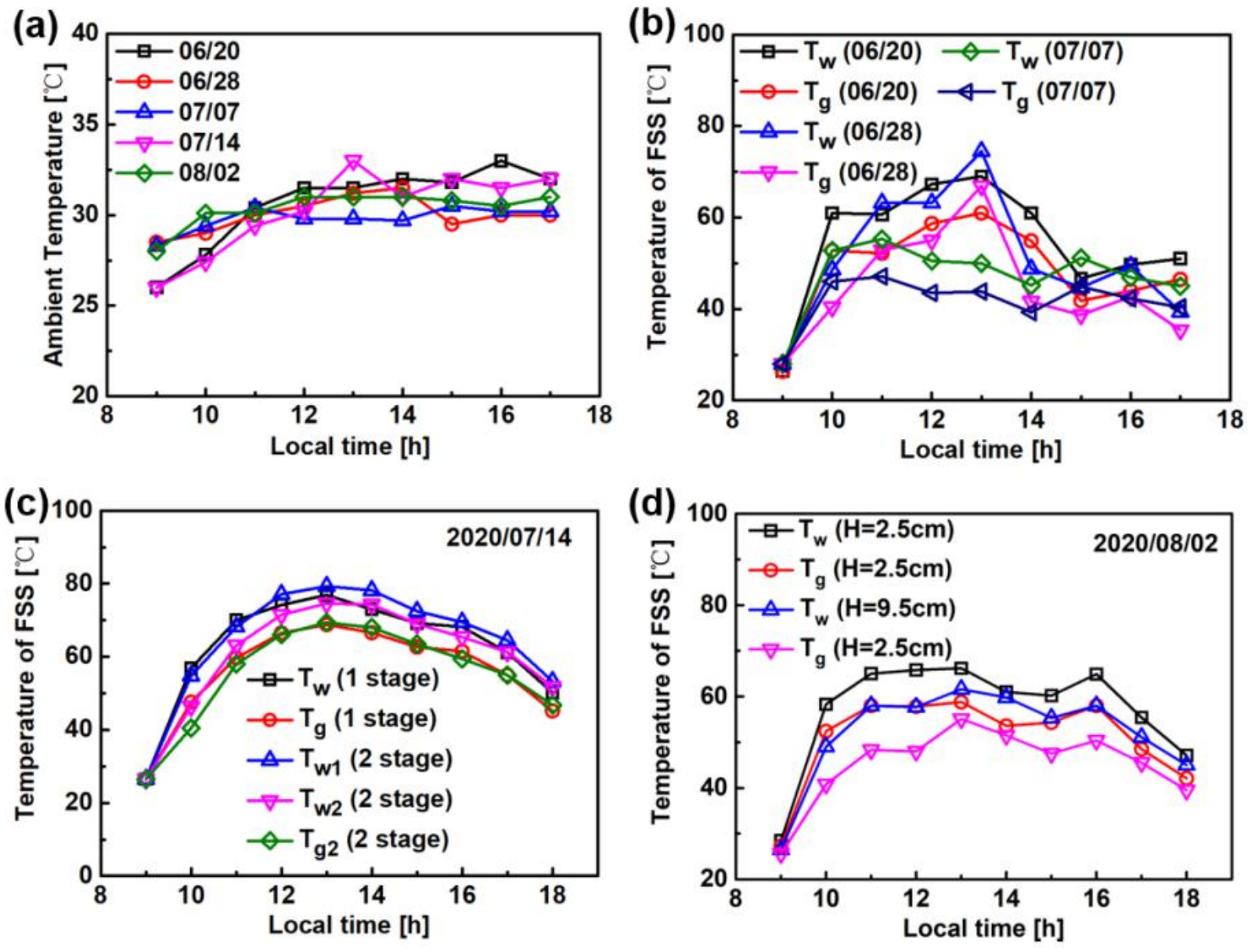

**Figure S11** Temperature of different days for ambient and solar still. (a) Ambient temperature of five different days. (b) Water temperature at the evaporation surface ($T_w$) and the temperature of glass cover ($T_g$) of FSS at different days. (c) $T_w$ and $T_g$ of single stage and double stage FSS. (d) $T_w$ and $T_g$ of FSS for H=2.5 cm and H=9.5 cm.

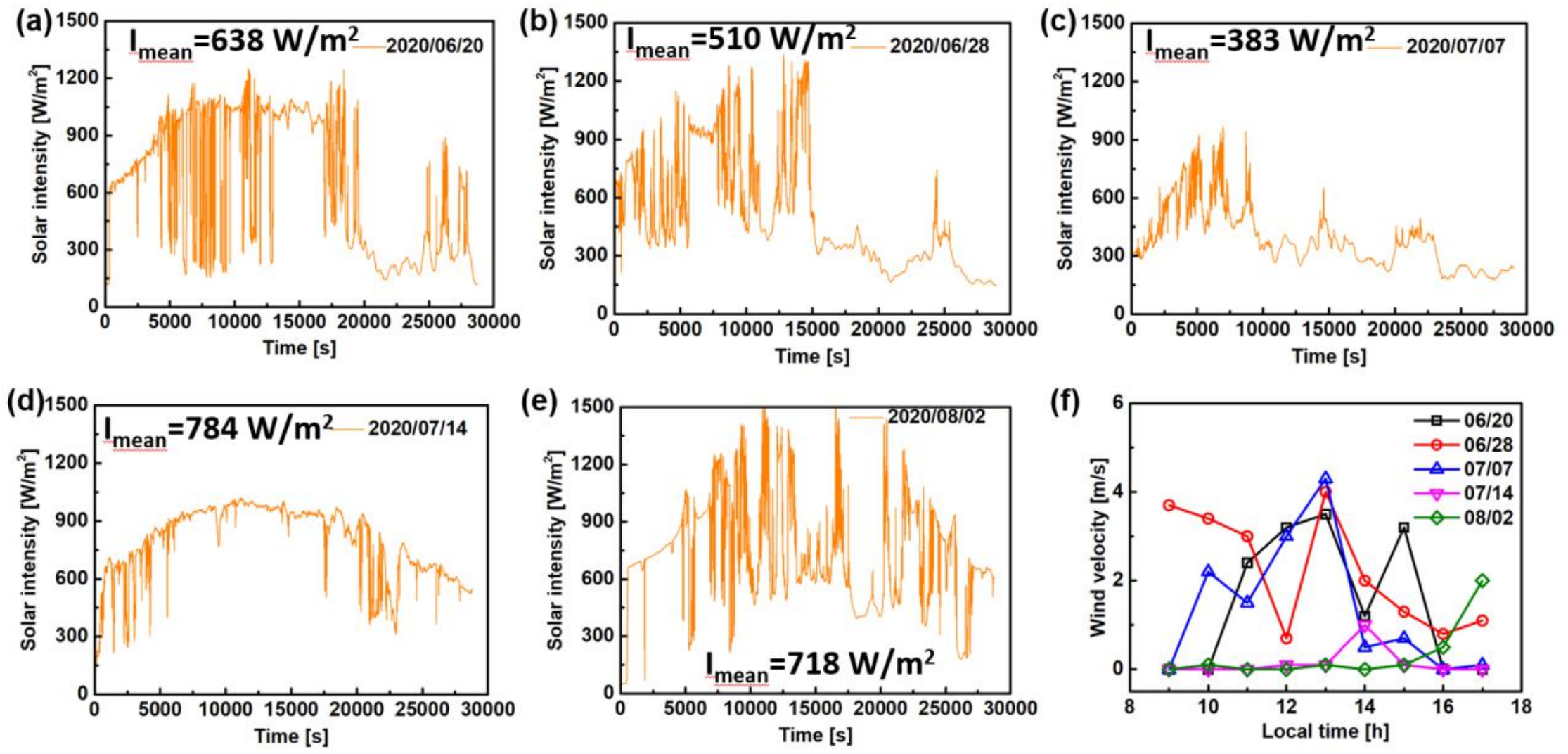

**Figure S12** Solar irradiation and wind velocity of different days. (a)-(e) Solar irradiation. (f) Wind velocity.